\documentclass[twocolumn,showpacs,prb,amsmath,amssymb,floatfix]{revtex4}

\usepackage{graphicx}
\usepackage{dcolumn}
\usepackage{bm}
\usepackage{latexsym}
\usepackage[usenames]{color}

\makeatletter
\def\lsim{\mathrel{\mathpalette\gls@align<}}
\def\gsim{\mathrel{\mathpalette\gls@align>}}
\def\gls@align#1#2{\lower.6ex\vbox
	{\baselineskip\z@skip\lineskip\z@\ialign
		{$\m@th#1\hfill##$\crcr#2\crcr\sim\crcr}}}
\makeatother
\def\be{\begin{equation}}
\def\ee{\end{equation}}
\def\bea{\begin{eqnarray}}
\def\eea{\end{eqnarray}}

\makeatletter
\def\leqq{\mathrel{\mathpalette\gle@align<}}
\def\geqq{\mathrel{\mathpalette\gle@align>}}
\def\gle@align#1#2{\lower.6ex\vbox{%
\baselineskip\z@skip\lineskip\z@\ialign{$\m@th#1\hfill##$\crcr#2\crcr=\crcr}}}
\makeatother

\begin{document}
\preprint{Two-Dimensional Nucleation with Edge and Corner Diffusion
}
\title{Two-Dimensional Nucleation with Edge and Corner Diffusion
}

\author{Yukio Saito}\email{yukio@rk.phys.keio.ac.jp}
\affiliation{
	Department of Physics, Keio University,
		3-14-1 Hiyoshi, Kohoku-ku, Yokohama 223-8522, Japan
}
\date{\today}


\begin{abstract}
The effect of edge  and corner diffusions on the morphology and
on the density of islands nucleated irreversibly on a flat substrate surface
is studied.
Without edge and corner diffusion, islands are fractal. As an edge
diffusion constant $D_e$ increases, islands tend to
take a cross shape with four needles in the $< 10 >$ direction.
Additional corner diffusion with a diffusion constant $D_c$ 
yields square islands. 
When $D_e$ is small relative to the surface diffusion constant $D_s$, 
the square corner shows the Berg instability to produce hopper growth in
the $<11>$ direction.
The corner diffusion influences the island number density $n$.
At a deposition flux $F$ with a small $D_c$,
mainly monomers are mobile and $n \propto (F/D_s)^{1/3}$.
At large $D_c$, dimers and trimers are also mobile and
$n \propto F^{3/7} D_s^{-5/21} D_c^{-4/21}$. The $F$ dependence
is in good agreement to the rate equation analysis, but the dependence
on $D_c$ cannot be explained by the theory.
\end{abstract}


\maketitle

\section{Introduction}

Surface morphology during epitaxial growth depends on processes 
taking place on the substrate.
\cite{hwang+91,michely+93,stroscio+93,brune+94,brune+95,hohage+96,li+97,brune+99}
At very low temperatures as in the case of molecular beam epitaxy (MBE), 
deposited atoms are adsorbed and migrate on the substrate surface
 until they form nuclei
or are incorporated into islands,
but they never evaporate back to the ambient vapor.
When an adsorbed atom (adatom, for short) sticks and 
freezes irreversibly to an island,
the resulting aggregate takes an irregular dendritic form,
\cite{hwang+91,michely+93,brune+94,brune+95,hohage+96} called 
diffusion-limited aggregate (DLA).\cite{witten+81}
It has a characteristic feature of self-similarity, 
and is regarded as a typical example of a fractal.
In actual experiments, one does not always observe fractal but also 
compact islands.\cite{michely+93,stroscio+93}
Compact islands are possible if they can dissociate
\cite{irisawa+95}
or if atoms at the edge of them can diffuse along 
the periphery.
\cite{michely+93,bales+94,zhang+94,bales+95,brune+99,ovesson+99,amar99,zhong+01}
Since at low temperatures it is difficult for an edge atom
to detach, the edge diffusion mainly governs the island morphology.
\cite{brune+99}
During the edge diffusion, however, 
the outer corner provides an additional barrier to be surpassed.
There, an edge atom  has to pass an intermediate state 
where the number of neighboring bonds is less than that
on the straight step edge. 
The edge and the corner diffusion is known to influence island morphology.
\cite{michely+93,bales+94,zhang+94,bales+95,hohage+96,li+97,amar99,brune+99,ovesson+99,zhong+01}
Without edge and corner diffusion, islands are similar to DLA. 
With the edge diffusion they take cross shapes, 
with the corner diffusion in addition
they become square. 
At the coalescence, these peripheral diffusion governs the shape relaxation.
\cite{stoldt+98}
Furthermore, the corner diffusion barrier is found to induce 
mound formation for multilayer deposition.
\cite{opl+99,murty+99}

Diffusion processes are also known to influence the number density $n$ of
islands nucleated on a singular surface.
\cite{venables73,stoyanov+81,venables+84,irisawa+90,villain+92a,villain+92b,tang93}
When the deposition rate $F$ is large, the adatom density
is high and the nucleation occurs frequently. Thus the island
density $n$ is expected to be high.
When an isolated adatom migrates on the surface with
a large surface diffusion constant $D_s$, 
the deposited atom moves
around a long distance, until it finds nucleation partners or 
preexisting islands and sticks to them. 
Thus, as the diffusion rate $D_s$ increases, 
the rate of forming new nuclei decreases, and 
the island density $n$ on a singular surface
diminishes.

More quantitatively, the island density $n$ is defined as the total 
number density
of immobile and undissociable clusters with sizes larger 
than the critical cluster size $i^*$.
In the classical mean-field nucleation theory,
the number density of adatom clusters of size $i$
is analyzed by means of rate equations.
\cite{venables73,stoyanov+81,villain+92a,villain+92b,furman+97}
When clusters smaller than $i^*$ can dissociate 
but only the mononer adatom is mobile with the diffusion constant $D_s$,
$n$ is found to follow the scaling law
\cite{venables73,stoyanov+81}
\begin{equation}
n \propto (\frac{F}{D_s})^{ i^*/(i^*+2) }.
\label{eq1}
\end{equation}
If the clusters with sizes $i$ smaller than the critical size $i^*$
cannot dissociate but are mobile as a whole with their respective diffusion 
constant $D_i$,
the scaling form is approximately given as 
\cite{villain+92a,villain+92b,furman+97}
\begin{equation}
n \propto (\frac{F^{i^*}}{D_s D_2 D_3 \cdots D_{i^*}})^{1/(2 i^*+1)}.
\label{eq2}
\end{equation}

By assuming that islands are point sinks to diffusing
adatoms and the critical nuclear size is $i^*=1$, 
the island density scaling 
$n \propto (F/D_s)^{1/3}$
is confirmed by the kinetic Monte Carlo (KMC) simulation.
\cite{bartelt+92}
 Actually, islands increase their size during growth, and
the scaling relation between the density and the diffusion constant
might depend on the island morphology. 
There are already many studies on the island density by means of 
KMC simulation when clustes can dissociate.
\cite{bales+94,liu+95,brune+99,mulheran+01}
There are also  KMC simulations with mobile clusters.
\cite{liu+95,bartelt+96,furman+97,mulheran+01}
We study here the effect of edge and corner diffusion processes on 
the island density and its relation to the island morphology, systematically.

In $\S$ 2, we explain our simulation scheme. In $\S$ 3, 
the effect of peripheral diffusion
on the island morphology is studied.
On varying the ratio between the surface, edge and corner diffusion,
morphology varies as fractal, needle, and square shapes, as 
 previously observed.
\cite{zhong+01}
We point out the different roles of the edge and corner diffusions 
on the island
symmetry. 
The island density is discussed in $\S$ 4.
As long as the corner diffusion is small, 
the density scaling is found to be almost independent of the island morphology 
with an exponent 1/3, in consistent to the more
systematic study by Brune et al.
\cite{brune+99}
With a large corner diffusion constant, on the other hand,
the scaling form of the island density changes drastically, because
small clusters of sizes less than 3
can migrate on the surface, i.e. $i^*= 3$.
The density is well fitted to the scaling $n \sim F^{3/7}$.
The increase of the critical island size $i^*$ is also reflected
in the island size distribution.
\cite{bartelt+96,amar+95,furman+97}
The result is summarized in a scaling plot of the island density,
and discussed in the last $\S$ 5.
Even though the scaling of the island density to the deposition flux
agrees to the rate equation analysis, the
dependence on the diffusion constants fails to follow the expectation
(\ref{eq2}).

\section{Monte Carlo simulation}

Our main aim is to explore the systematics and the universal features
of the morphology change of islands on a singular surface 
as the diffusion mechanism changes, rather than the precise reproduction 
of some specific experimental observations.
For this purpose, we adopt the simplest model such that atoms 
are depositing on  a (100) surface of a simple cubic lattice 
with a solid-on-solid (SOS) restriction. 
The SOS restriction is valid at low temperatures, common to MBE.

Atoms are deposited randomly onto the substrate with a deposition rate of $F$
monolayers per unit time. Here we choose the unit of time such that $F=1$.
The temperature is assumed to be so low that all deposited atoms stick 
to the surface and never evaporate back into the ambient gas.
An isolated adatom hops to one of its four nearest
neighboring sites at a rate $k_s$; the waiting time of an adatom before
the jump is $(4k_s)^{-1}$, and the surface diffusion constant is given
by $D_s=k_s a^2$, where $a$ is the lattice constant.
As for the interlayer diffusion, which takes place in case a deposited atom
lands on another adatom, no additional energy barrier is assumed for it
to diffuse down.
On the contrary, atoms at the edge are not allowed to hop up a layer.
Therefore, Ehrlich-Schwoebel mound formation is supressed, and
the layer-by-layer growth is expected.

\begin{figure}[t]
\begin{center} 
\includegraphics[width=0.95\linewidth]{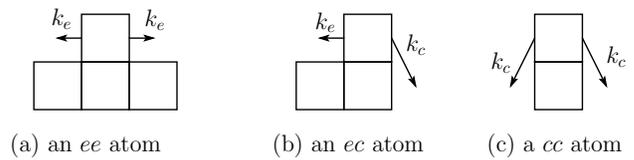}
\end{center} 
\caption{
Diffusion processes of an adatom connected to an island with a single
in-layer bond. $k_e$ is the rate of edge diffusion and $k_c$ is that of
corner diffusion.}
\label{fig1}
\end{figure}

When an atom touches to other adatoms or islands, it makes a nearest neighbor
 bond to form a cluster.
At very low temperatures, this cluster will never dissociate again. 
We assume that adatoms attach clusters irreversibly, and no cluster
dissociation takes place.
This means that dimers are stable, 
and the critical island size is expected to be $i^*=1$.
Nevertherless, if an edge atom is singly-bonded in the layer, i.e. 
if it makes only one bond 
with another adatom, it is possible to diffuse along the island periphery at
moderate temperatures.
There are two possible diffusion jumps for a singly-bonded 
 atom; an edge diffusion along the straight step edge, and a corner
diffusion or corner crossing round the outer corner.
The rate of edge diffusion is $k_e$ and that of corner is $k_c$, as
is depicted in Fig.\ref{fig1}. 
Then, a singly-bonded atom can be classified by 
its possible motions in two directions.
It is classified as (a) an $ee$ atom on a straight step 
if both movements are along the step edge,
(b) an $ec$ atom at the ridge of a corner 
if one is the edge diffusion and the other is the corner,
and (c) a $cc$ atom on a tip if both movements are across the corner, 
as shown in 
Fig.\ref{fig1}.
The edge diffusion constant along the straight step edge is $D_e =k_e a^2$,
and we may define the corner diffusion constant as $D_c=k_c a^2$.
Precise values of diffusion constants depend on the energy barriers and
the temperature.
For edge and corner diffusion there may exist extra energy barriers 
in addition to that for the surface diffusion,
and  diffusion constants are probably ordered as $D_s \ge D_e \ge D_c \ge 0$.
If an edge atom is incoorporated into the kink site with more than two 
nearest neighbors,
it is assumed to cease migration, for simplicity.

Simulation starts from a clean substrate surface of a size $L \times L$
without adatoms. Hereafter we take the lattice constant $a=1$.
Therefore, the jump rates $k$'s and diffusion constants $D$'s are the same.
During the simulation, at some stage, there are $N_0$
isolated adatoms, $N_{ee}$ edge atoms attached to a straight step edge,
$N_{ec}$ atoms at the corner, and 
$N_{cc}$ atoms at the tip position (see Fig.\ref{fig1}) on the surface.
These adatoms are sorted out in corresponding lists.
The transition probability from this configuration to
the next one is given as follows.
The rate of the deposition is $P_d=F L^2$,
of the  surface diffusion $P_s=4 N_0 D_s $,
of the edge diffusion  $P_e=(2 N_{ee}+N_{ec}) D_e$ and 
of the corner diffusion  $P_c=(N_{ec}+2 N_{cc}) D_c$.
The total probability of a state change in a unit of time is
$P_t=P_d+P_s+P_e+P_c$. Thus, within a time interval $dt=1/P_t$,
one of the events takes place; the deposition with a probability
$P_d dt$, the surface diffusion with $P_s dt$, the
edge diffusion with $P_e dt$ and the
corner diffusion with $P_c dt$.
For each diffusion process, the atom to be moved is picked up 
randomly from the 
list. After each state change, adatom lists have to be adjusted, for example,
by eliminating those adatoms with more than two neighboring bonds
from the lists, or add an adatom that happens to have a single bond 
to the corresponding list.
The time is increased by $dt$. 
Since the time unit is chosen such that the deposition flux $F=1$, 
the monolayer is covered at a time $t=1$,
in principle. 
Because of the stochastic feature of the simulation algorithm, in practice,
the actual time for the completion of a monolayer fluctuates around $t=1.$

\section{Island Morphology}

\begin{figure}[h]
\includegraphics[width=0.32\linewidth]{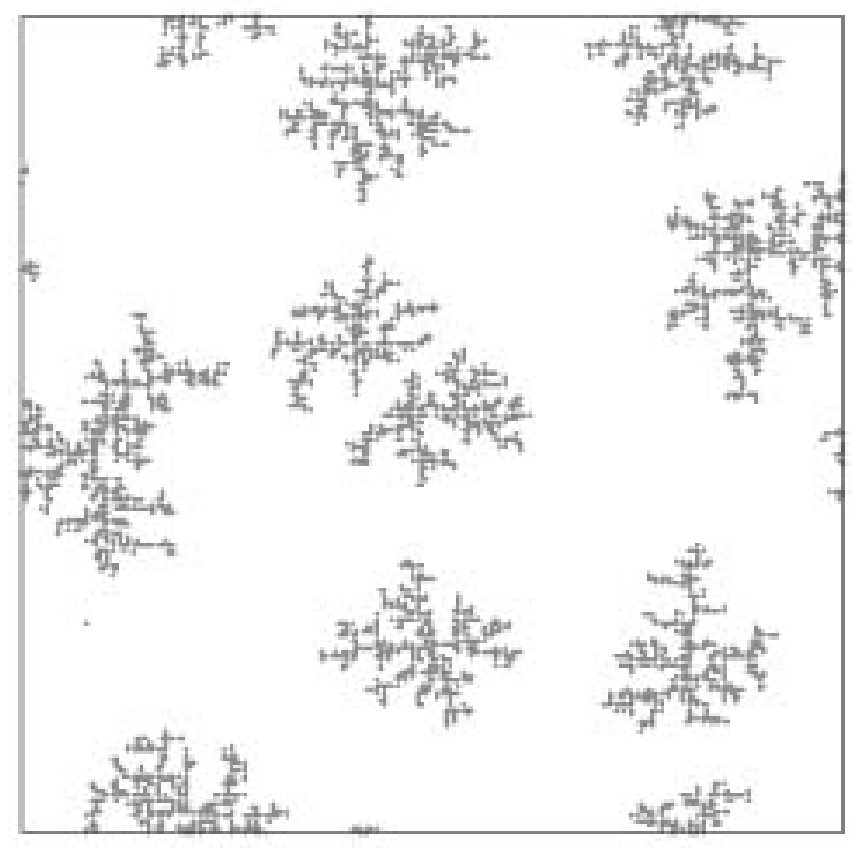}
\includegraphics[width=0.32\linewidth]{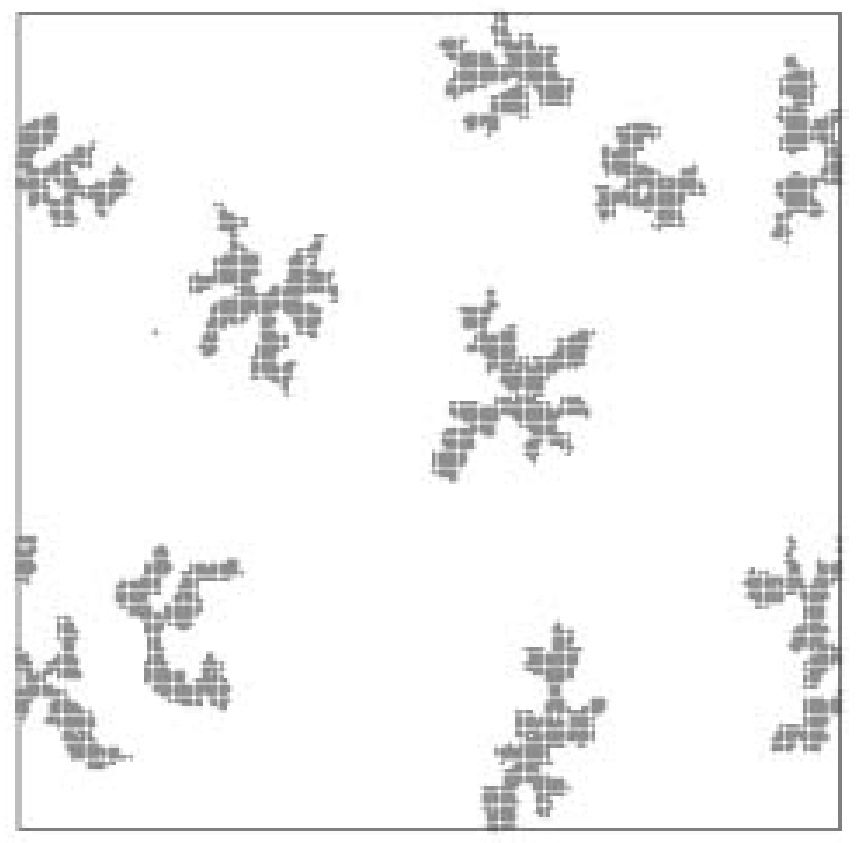}
\includegraphics[width=0.32\linewidth]{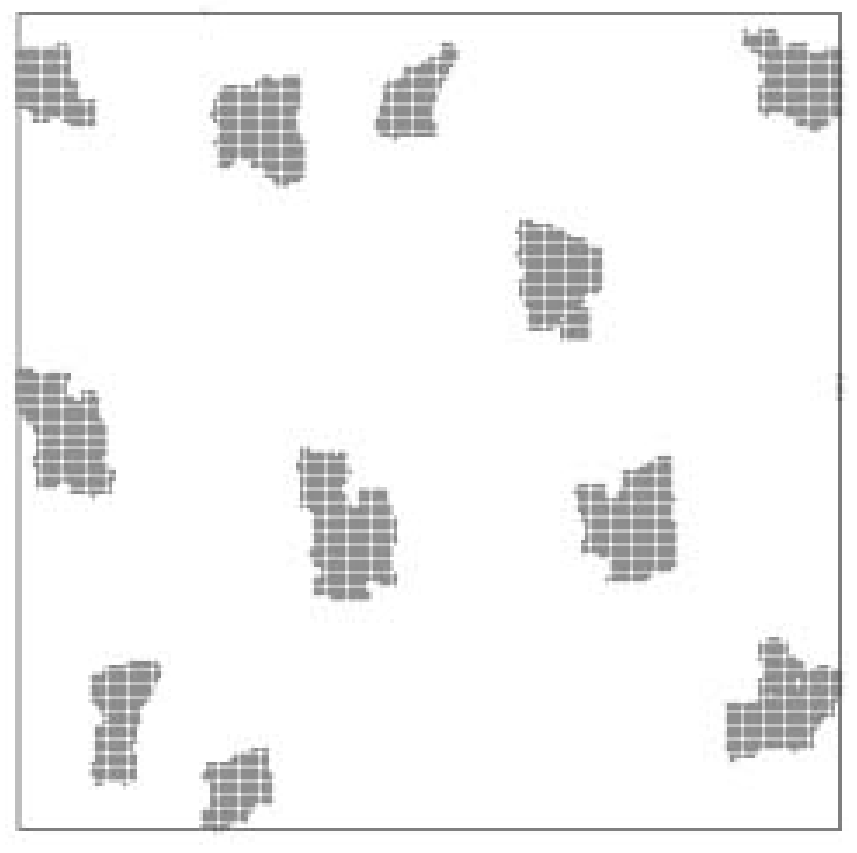}
\\ \hspace*{.5cm}(a) (0, 0) \hspace{0.8cm}(b) ($10^{-6}$, $10^{-6}$)
  \hspace{.2cm}(c) ($10^{-5}$, $10^{-5}$)\\
\includegraphics[width=0.32\linewidth]{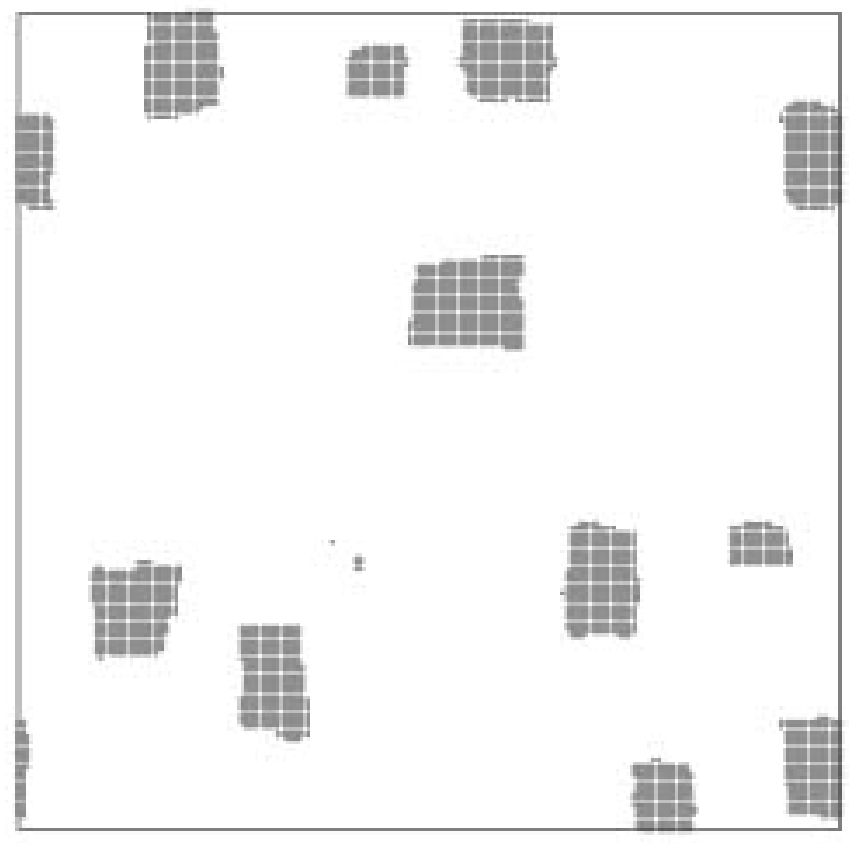}
\includegraphics[width=0.32\linewidth]{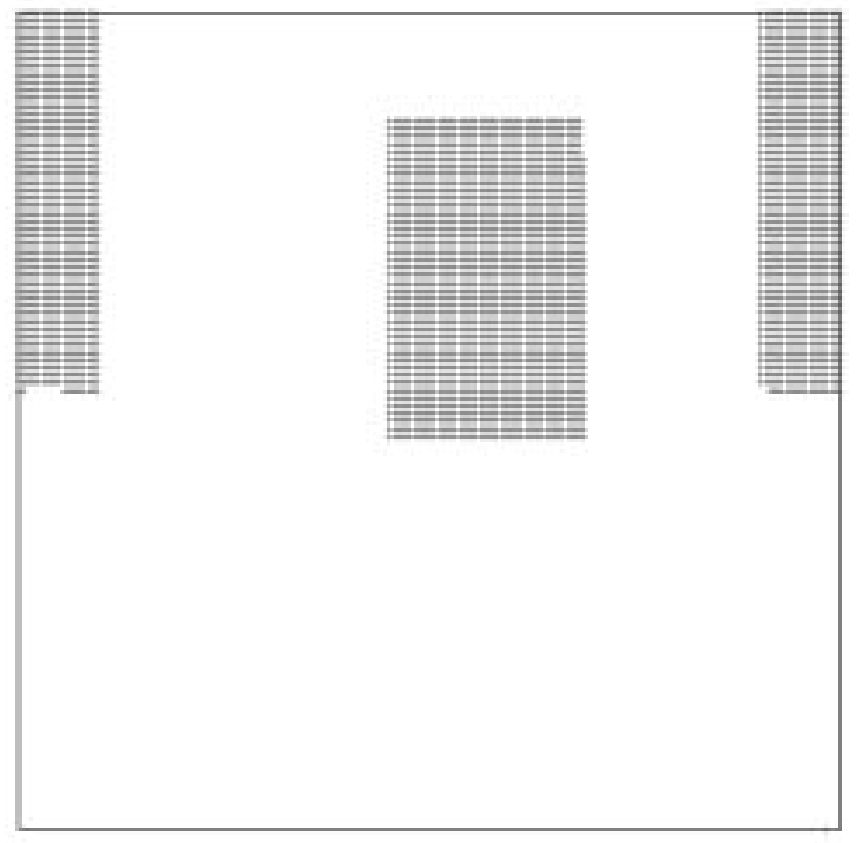}
\includegraphics[width=0.32\linewidth]{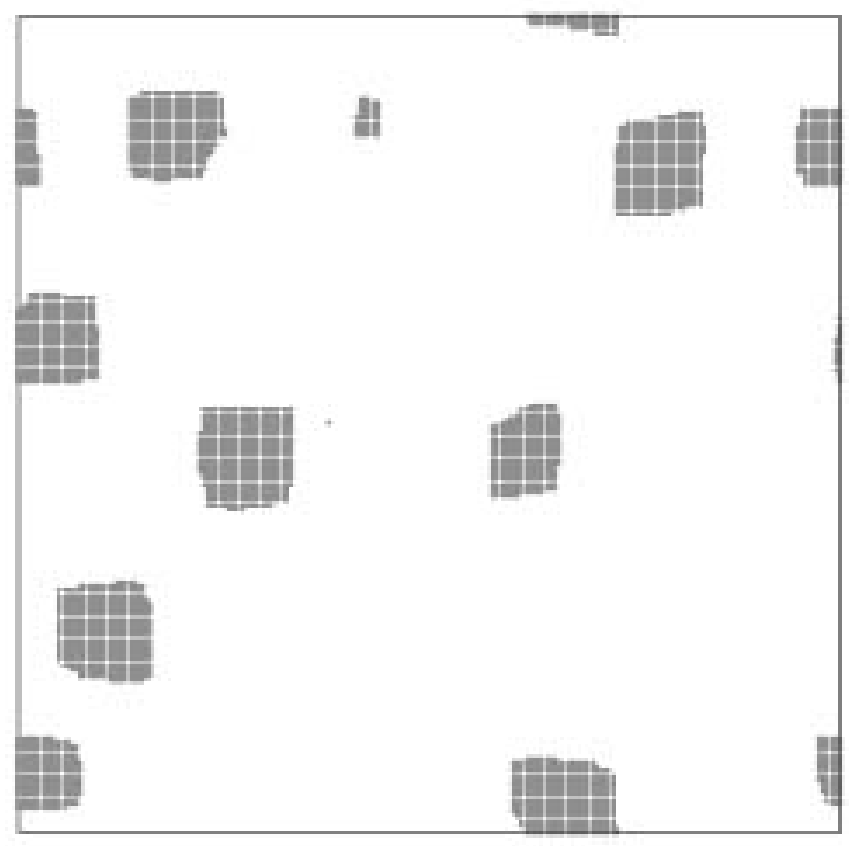}
\\ (d)  ($10^{-4}$, $10^{-4}$) \hspace{.5cm}(e)  (1, 1) 
\hspace{1.5cm}(f)  (1, $10^{-4}$)\\
\includegraphics[width=0.32\linewidth]{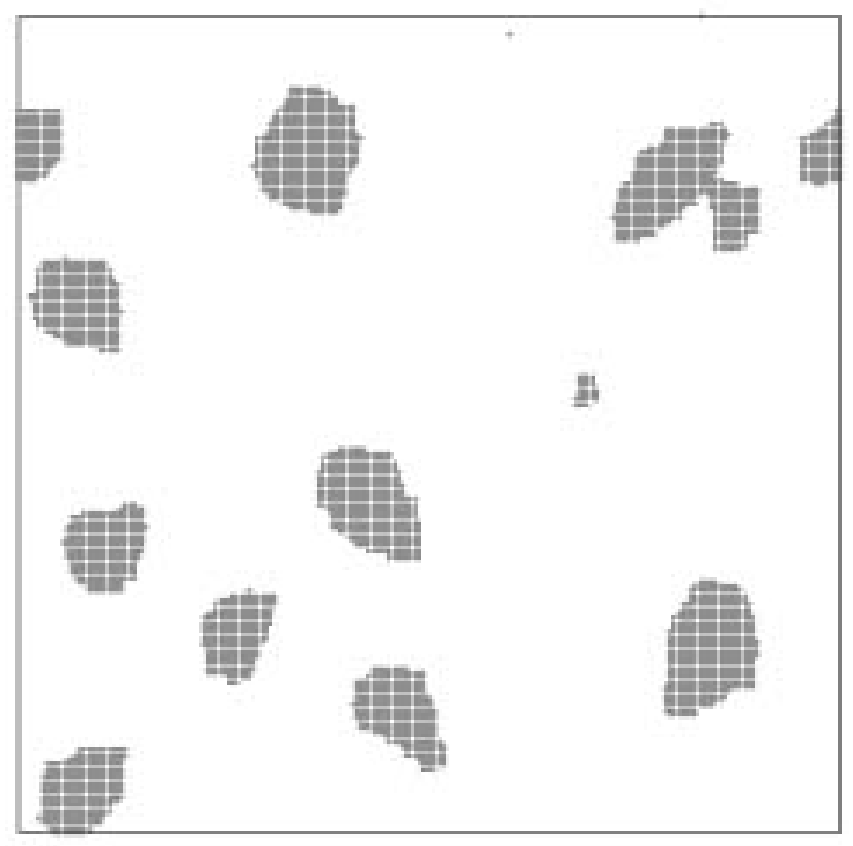}
\includegraphics[width=0.32\linewidth]{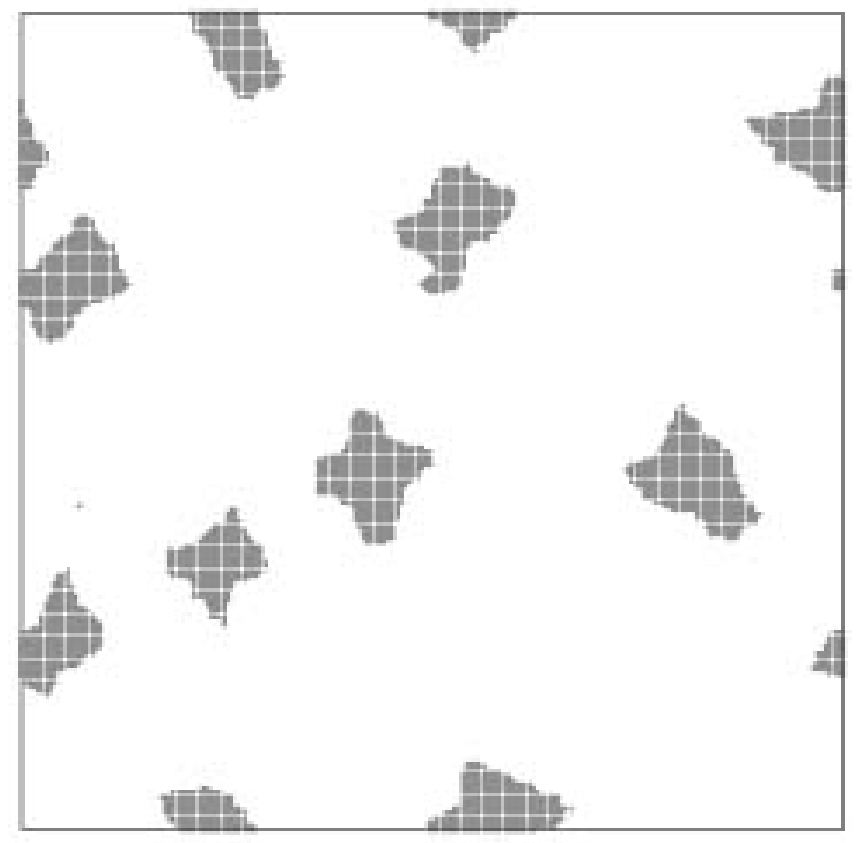}
\includegraphics[width=0.32\linewidth]{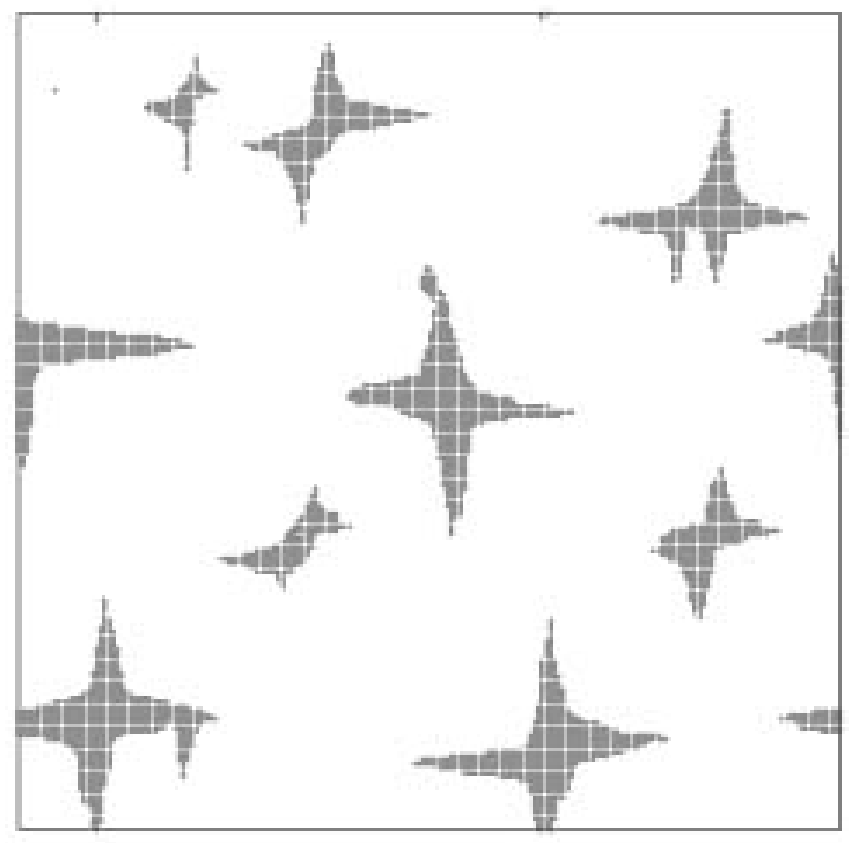}
\\(g) (1, $10^{-5}$) \hspace{0.8cm} (h) (1, $10^{-6}$)  
\hspace{1.cm}(i)  (1, 0)\\
\includegraphics[width=0.32\linewidth]{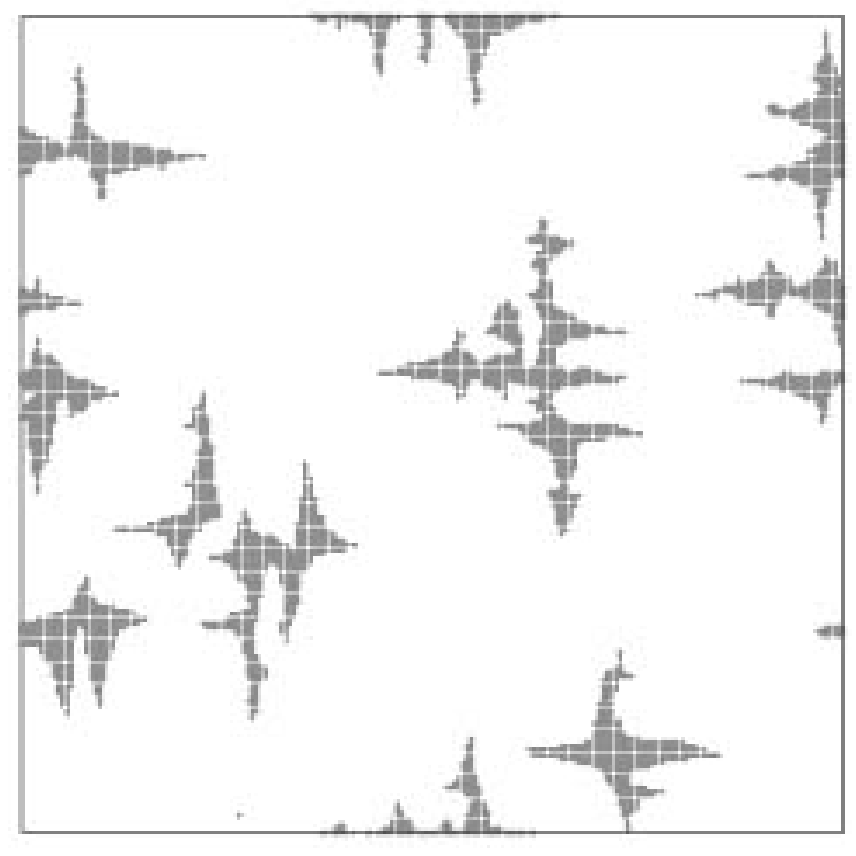}
\includegraphics[width=0.32\linewidth]{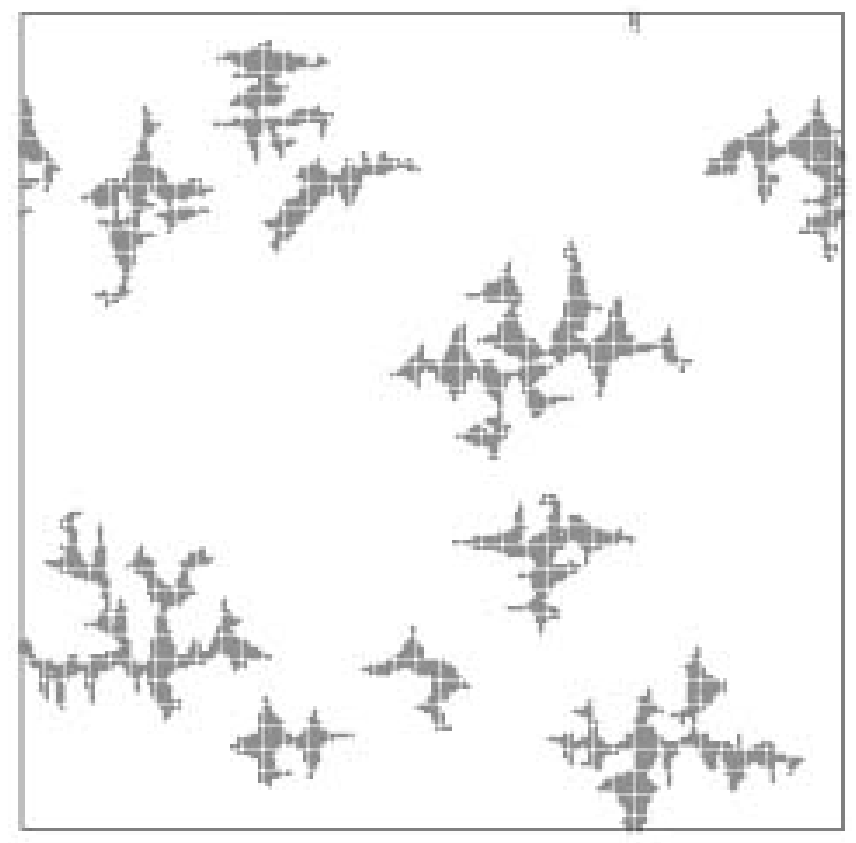}
\includegraphics[width=0.32\linewidth]{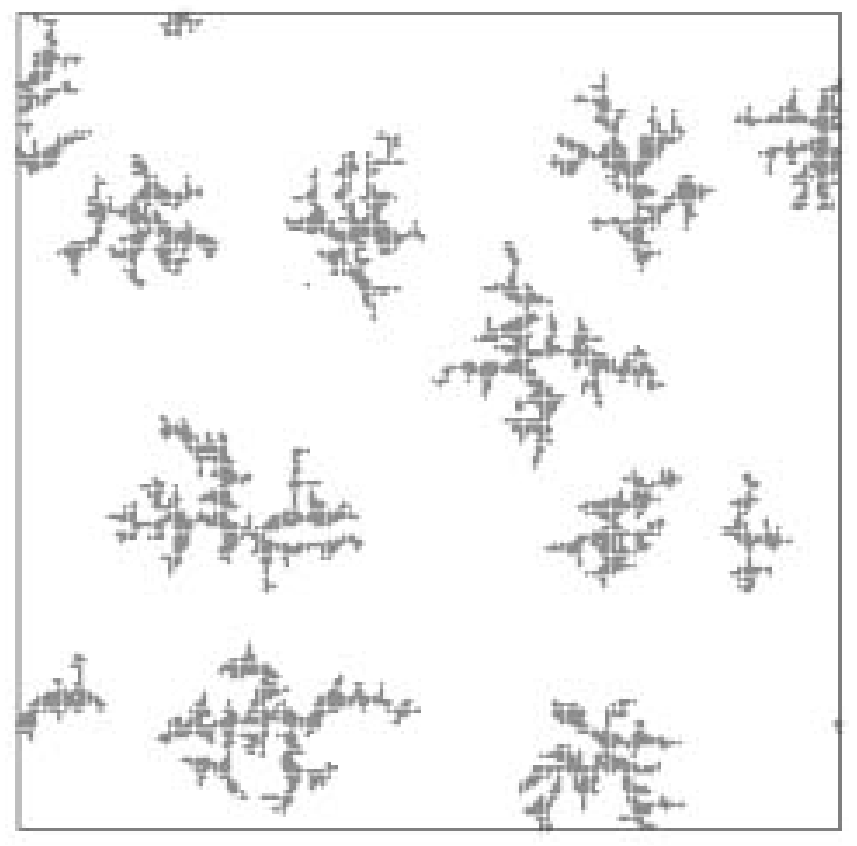}
\\ (j) ($10^{-4}$, 0)\hspace{1.cm}(k) ($10^{-5}$, 0) 
\hspace{1.cm}(l) ($10^{-6}$, 0)
\caption{
Island morphology at various combinations of the surface, edge and corner 
diffusion constants, $D_s$,$D_e$ and $D_c$. 
$D_s$ is fixed at $D_s/F=10^9$. 
Parameters in parenthes represents $(D_e/D_s, D_c/D_s)$.
The system size is 200$\times$200, and
the coverage is $\theta=0.1$ML.
}
\label{fig2}
\end{figure}

Simulations are performed for systems of sizes $200 \times 200$,
and $1000 \times 1000$ under various combinations of diffusion constants,
and some typical island morphology is shown in Fig.\ref{fig2} at a coverage 
of $\theta=$0.1 monolayer(ML).

From Fig.\ref{fig2}(a) to (e) the edge and corner diffusion constants, 
$D_e$ and $D_c$, 
are equally increased from 0 to $D_s$, by keeping the surface diffusion 
constant $D_s$ fixed at $D_s/F=10^9$.
Without peripheral diffusion, the island takes an irregular dendritic
form or a fractal DLA-like shape, as shown in Fig.\ref{fig2}(a). 
On increasing the edge and corner diffusion to $D_e=D_c=10^{-6} D_s$,
 dendrite branchs thicken and they extend mainly in the $<11>$ direction, as
 shown in Fig.\ref{fig2}(b).
Further increase of $D_e=D_c=10^{-5}D_s$ 
let the islands compact in a hopper shape with a corner instability due to
Berg effect,\cite{berg38} as in Fig.\ref{fig2}(c).
Since the corner is extending in the region with high monomer concentration,
the corner captures many adatoms and the edge diffusion transfers them
to the island center. But the edge diffusion is still weak and 
insufficient, and the growth of the central part is too slow to catch up the 
fast corner growth.
Thus, the  island grows mainly in the $<11>$ direction, 
as in Fig.\ref{fig2}(b) and (c).
With sufficient edge and corner diffusion as $D_e=D_c=10^{-4}D_s$, 
islands take square shape surrounded by rather straight 
\{10\} steps, as in Fig.\ref{fig2}(d).
So far island density seems independent on the island morphology.
When all the diffusion constants are equal, as in Fig.\ref{fig2}(e), 
islands are
square and very large, but their density decreases drastically.
On this density change we shall discuss later in \S 4.
The sequence of simulations from Fig.\ref{fig2}(b) to (d) almost corresponds to 
the parameter range studied by Bales and Chrzan\cite{bales+94} by changing the
temperature. The morphology variation is of course quite analogous to
what they have obtained.

We now study the effect of corner diffusion more precisely, in a sequence of
pictures shown from Fig.\ref{fig2}(e) to (i).
There, the edge diffusion constant is
kept as large as $D_e=D_s$, and only the corner diffusion constant $D_c$ is
decreased from $D_s$ to 0. In all these cases, islands are not irregular.
With a large corner diffusion constant, $D_c \ge 10^{-4} D_s$,
island are square with \{10\} steps, as shown in Fig.\ref{fig2}(e) and (f).
 At $D_c=10^{-5} D_s$ in Fig.\ref{fig2}(g),
they look round and rather isotropic, and for a still smaller $D_c =
10^{-6}D_s$ in Fig.\ref{fig2}(h) some islands are in diamond shape and 
some others are in a cross shape with thick arms.
In the extreme case of Fig.\ref{fig2}(i) 
without the corner diffusion, $D_c=0$, islands consist four sharp needles
growing in the $<10>$ direction.
The same four-needle shape is observed previously by Zhong et al. 
\cite{zhong+01}
in a submonolayer growth and by Caspersen et al.\cite{caspersen+01}
 in a multilayer study.

In a series of simulations  Fig.\ref{fig2}(i)-(l) 
the corner diffusion is completely
suppressed ($D_c=0$), and the effect of the edge diffusion $D_e$ is
 exemplified.
The series, in fact, continues back to  Fig.\ref{fig2}(a).
With $D_e=D_s$ in  Fig.\ref{fig2}(i) 
needles in cross-shape islands are rather stable 
and sidebranches are rare.
On decreasing the edge diffusion to $D_e =10^{-4}D_s$ in  Fig.\ref{fig2}(j), 
needles have
many sidebranches growing in the $<10>$ direction. 
Further decrement to $D_e =10^{-5} D_s$ leads to
tip splitting in Fig.2(k). With still smaller $D_e =10^{-6}D_s$
clusters look like DLA aggregates. 
Comparison with  Fig.\ref{fig2}(b) and (l)
reveals that the corner diffusion produces $<11>$ dendrite in  Fig.\ref{fig2}(b),
whereas without corner diffusion aggregates have a preference in 
the $<10>$ direction
in  Fig.\ref{fig2}(l).
Also without corner diffusion, the dendrite tip is pointed, whereas
with a corner diffusion, the tip becomes fat.
At $D_e=D_c=0$, the DLA looks isotropic. 
Of course, it is known to be anisotropic asymptotically by 
reflecting the underlying square symmetry of the lattice, 
but the shape anisotropy
appears only  when the DLA grows very large.
In the present simulation, islands coagulate to acomplish a layer growth
before the asymptotic anisotropy appears.

From these variety of island morphology the edge and corner diffusions
are found to play different roles on the island symmetry. 
The edge diffusion smears out shot noise
introduced by the irreversible attachment, and permits
the formation of regular shape. 
The corner diffusion varies the orientational preference.
Without the corner diffusion, those atoms on the tip positions or on the
narrow step edges cannot cross to the other sides of an island, 
and thus the supersaturation on the narrow \{10\} side increases. 
This leads to the enhancement in the growth
velocity normal to the narrow \{10\} side, and due to the strongly 
anisotropic kinetic effect a needle tip
is  eventually formed into the $<10>$ direction.
A corner diffusion allows the transfer of adatoms from one step edge to 
the other, and islands tend to be square. 
If the edge diffusion is weak, however, the adatom supply at 
the center of the step edge is meager, 
and the corner instability takes place.
This leads to the dendrite growth in the diagonal $<11>$ direction, as in 
 Fig.\ref{fig2}(b) and (c).

\section{Island Density}

In  Fig.\ref{fig2} at a fixed surface diffusion constant $D_s$, we observe that 
the island density is almost independent of the edge and corer diffusion
 constants, $D_e$ and $D_c$. 
 We said "almost", because in  Fig.\ref{fig2}(e) the island
density is clearly different from the others.
We now study the effect of peripheral diffusion on the island density.

The rate equation analysis predicts that
the island density $n$ depends on the surface diffusion constant 
as in eqs.(\ref{eq1}) and (\ref{eq2}).
\cite{venables73,stoyanov+81,villain+92a,villain+92b}
Experimentally, temperature dependence of the island density
is used to estimate values of various energy barriers by assuming
the critical island size $i^*$.
\cite{venables+84}
Simulations where the temperature is varied also has to
assume $i^*$ to interpret their results.
\cite{venables+84}
To study the relation of island morphology and its density by fixing $i^*=1$, 
we control directly on the rate of edge and corner diffusion, rather
than tune many parameters as bond energies, energy barriers and the
temperature.

\begin{figure}
\begin{center} 
\includegraphics[width=0.98\linewidth]{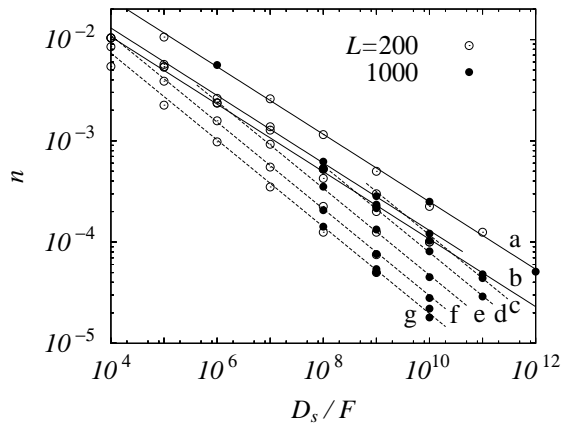}
\end{center} 
\caption{
Double logarithmic plot of the island density $n$ at the coverage 
$\theta=0.1$ML
versus the surface diffusion 
constant normalized by the deposition rate $D_s/F$.
The values of edge and corner
diffusion constants, $D_e$ and $D_c$, are 
(a) $D_e=D_s$ and $D_c=0$ (no corner diffusion), 
(b) $D_e=D_c=0$ (without edge and corner diffusion), 
(c) $D_e=D_c=10^{-4} D_s$, 
(d) $D_e=D_c=10^{-3} D_s$, 
(e) $D_e=D_c=10^{-2} D_s$, 
(f) $D_e=D_c=10^{-1} D_s$, and 
(g) $D_e=D_c=D_s$.
In (a) the horizontal axis is displaced to the right by a factor 10,
for a visual purpose.
Lines represent scaling behaviors as $n \propto (D_s/F)^{-\chi}$.
An exponent $\chi$ for straight lines is $\chi=1/3$, and for dashed lines 
$\chi=3/7$.
}
\label{fig3}
\end{figure}

The island density $n$ at the coverage $\theta=$0.1ML is plotted in Fig.\ref{fig3}
as a function of the surface diffusion constant $D_s/F$  
for various combinations of  the edge and corner diffusion constants.
The system sizes plotted are $200^2$ (open circles) 
and $1000^2$ (filled circles).

For the case without edge and corner diffusion, $D_e=D_c=0$, 
islands are fractal as shown in  Fig.\ref{fig2}(a), at least at low densities 
with a large $D_s$.
The island density is shown in Fig.\ref{fig3}(b).
Data fits well with the scaling relation $n =0.23 (D_s/F)^{-1/3}$.
The result is consistent to that obtained by Venables et al.
\cite{venables+84} $n=0.25 (D_s/F)^{-1/3}$ from the rate equation analysis
and that by
Brune et al. 
\cite{brune+99}
 $n =0.27 (D_s/F)^{-1.027/3}$
for the DLA cluster in kMC simulation at the coverage 0.12ML.
For the DLA, in fact, another scaling law 
$n \propto (D_s/F)^{-2/(4+D_f)}$ is proposed,
where $D_f=1.71$ is the fractal dimension.
\cite{villain+92a,villain+92b,tang93}
From the present simulations of sizes up to $1000^2$, 
we cannot descriminate which of the two scaling laws is superior.

For the case with a large edge diffusion $D_e=D_s$  
but without corner diffusion $D_c=0$,
islands have cross shapes as in  Fig.\ref{fig2}(i). 
Even for this situation, the density $n$
is compatible with a scaling law $n = 0.25 (D_s/F)^{-1/3}$,
as is shown in Fig.\ref{fig3}(a). The plot is shifted by a factor 10 to the right
for the visual purpose. Otherwise, data overlaps almost completely
with those for fractal islands, Fig.\ref{fig3}(b).

\begin{figure}
\begin{center}
\includegraphics[width=0.45\linewidth]{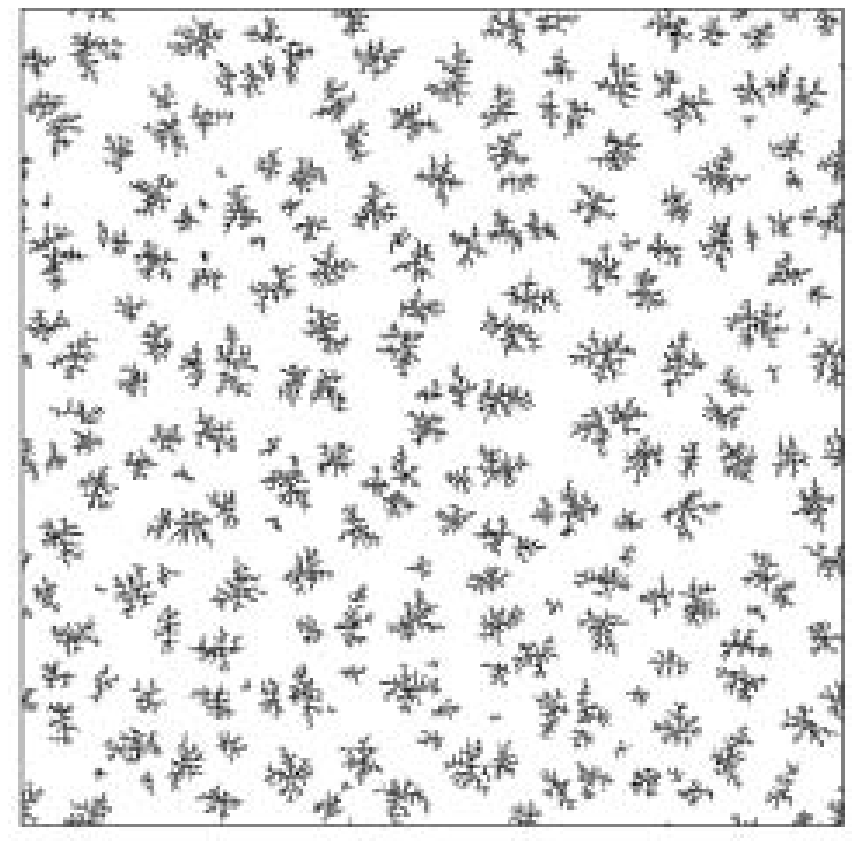}
\includegraphics[width=0.45\linewidth]{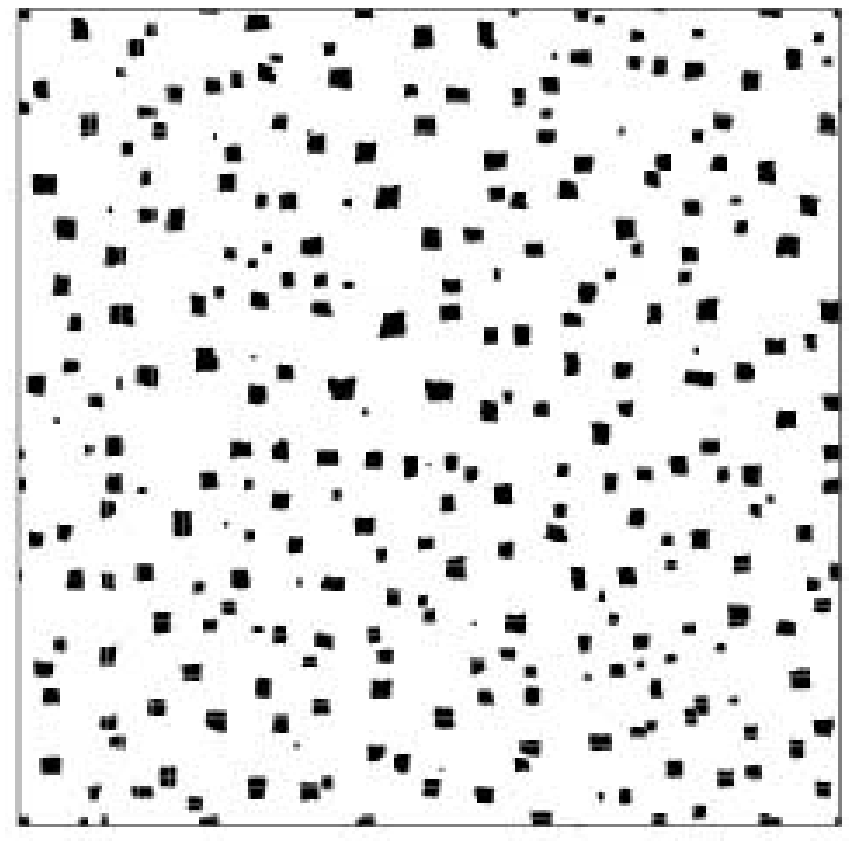}\\
(a)  \hspace{3.8cm} (b)  \\
\includegraphics[width=0.45\linewidth]{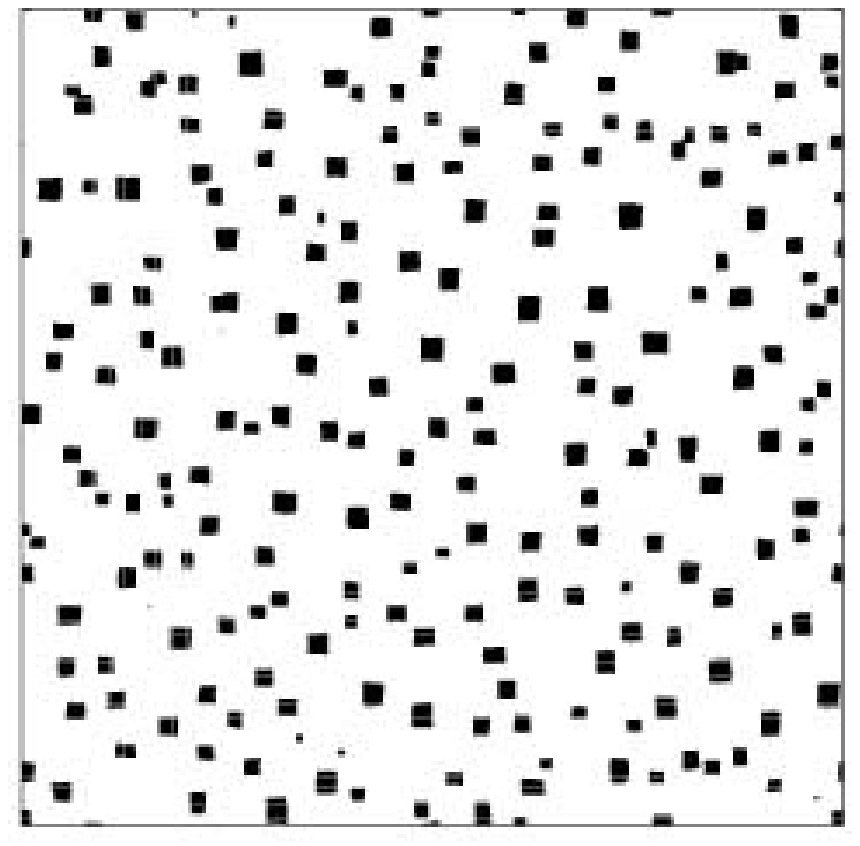}
\includegraphics[width=0.45\linewidth]{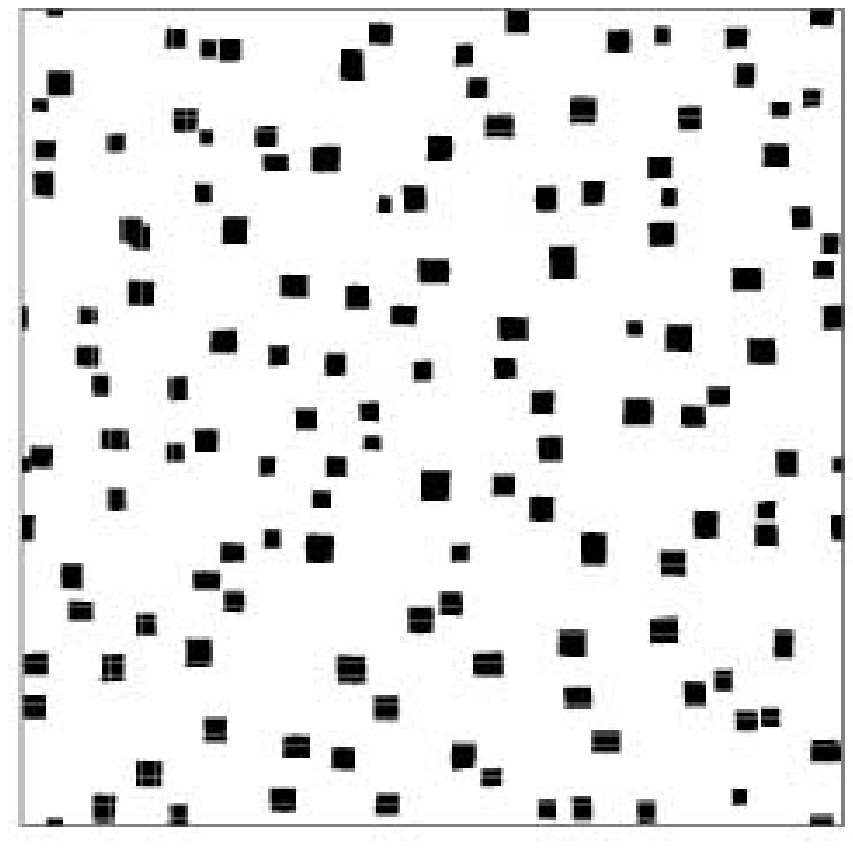}\\
(c) \hspace{3.8cm} (c)\\
\includegraphics[width=0.45\linewidth]{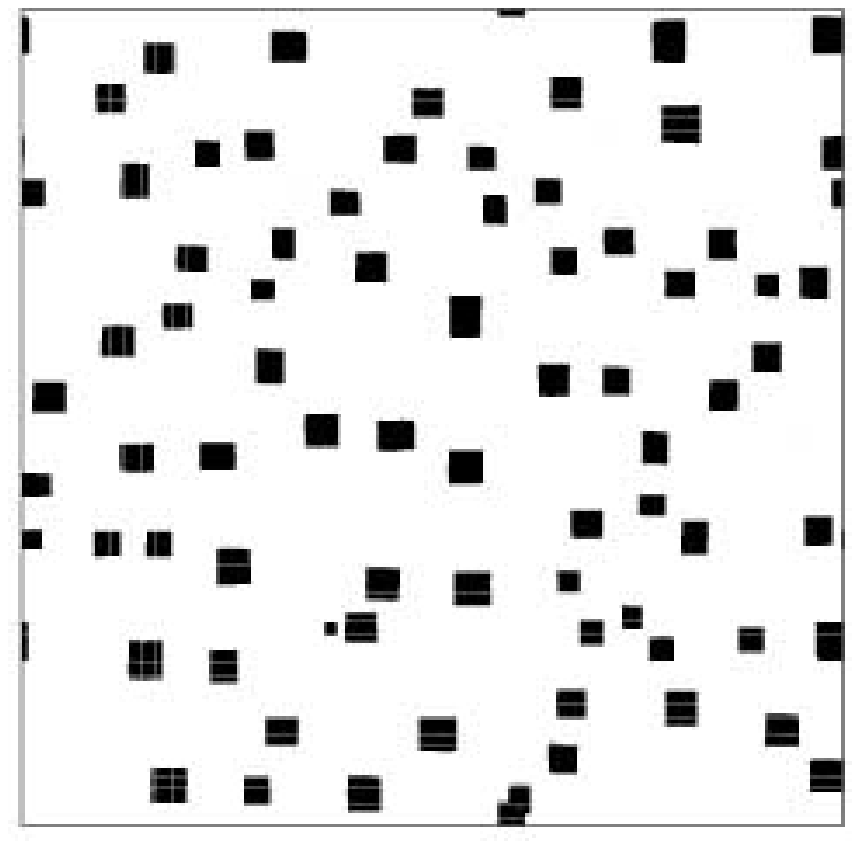}
\includegraphics[width=0.45\linewidth]{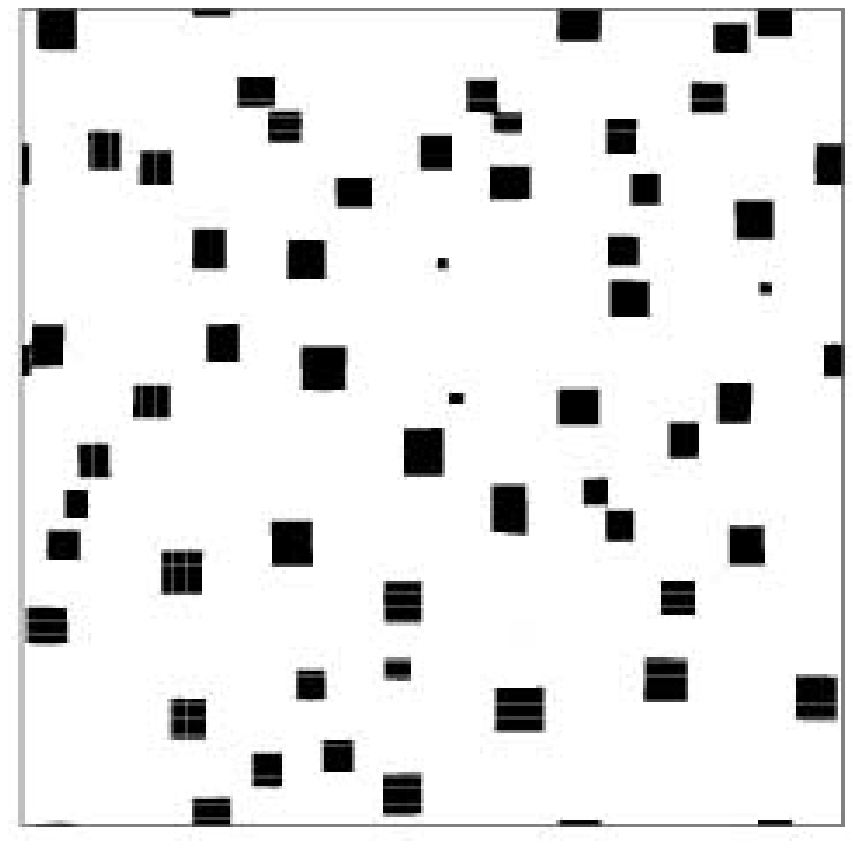}\\
(e) \hspace{3.8cm} (f)\\
\end{center}
\caption{Islands on a substrate of a size 1000$^2$ at a coverage $\theta=0.1$ML.
$D_s/F=10^9$ and $D_e=D_c$.
The  ratio of the corner to surface diffusion constant $D_c/D_s$
is (a) 0, (b) $10^{-4}$, (c) $10^{-3}$, (d) $10^{-2}$, 
(e) $10^{-1}$, and (f) 1.
}
\label{fig4}
\end{figure}

With or without edge diffusion, the island density is found to follow the 
same scaling relation with the surface diffusion constant $D_s/F$, obtained
by a simple rate equation theory.\cite{venables73,stoyanov+81}
Inclusion of the corner diffusion, however, drastically changes the situation. 
By keeping $D_e = D_c$ but at different ratio to the surface diffusion as
(c) $D_e/D_s=D_c/D_s=10^{-4}$, (d) $10^{-3}$, (e) $10^{-2}$, (f) $10^{-1}$, 
and (g) 1,
the relation between the island density $n$ and the surface diffusion constant
$D_s/F$ varies as shown in Fig.\ref{fig3}(c)-(g).
For (c) the island density at low $D_s/F \le 10^{10}$ is higher
than that for (b) with fractal islands by about 20 percent, 
$n= 0.28 (D_s/F)^{-1/3}$.
The density increase is due to the compactness of the islands,
as shown in the surface configuration in Fig.\ref{fig4}(b) 
for $D_e/D_s=10^{-4}$, compared to 
the fractal islands in Fig.\ref{fig4}(a) for $D_c=0$.
There are more space for adatoms to nucleate new compact islands.
As for the $D_s$ dependence, the island density follows the same scaling 
law with an exponent 1/3, 
unless $D_s$ takes the largest value $D_s/F =10^{11}$.
There, the density becomes less than that of fractal islands, Fig.\ref{fig3}(b).

As $D_e$ and $D_c$ increase further from Fig.\ref{fig3}(d) to (g),
the island density starts to deviate from the curve Fig.\ref{fig3}(c)
at smaller $D_s/F$.
For the case Fig.\ref{fig3}(d) with $D_e/D_s=D_c/D_s=10^{-3}$,
deviation takes place at about $D_s/F=10^9$,
for (e) at about $D_s/F=10^7$. For (f) and (g) with larger $D_e=D_c$,
the island density remains always less than that of fractal islands with
the same $D_s$. 
This tendency of density decrease is obvious by plotting the
surface configurations in Fig.\ref{fig4}. There, the coverage $\theta=0.1$ML and the 
surface diffusion constant $D_s/F =10^9$ are the same, but the
edge and corner diffusion constants, $D_e/D_s = D_c/D_s$, are
varied. On increasing $D_c$ from Fig.\ref{fig4}(b) to (f), 
the number of islands
decreases and the island size increases.

\begin{figure}[h]
\begin{center} 
\includegraphics[width=0.8\linewidth]{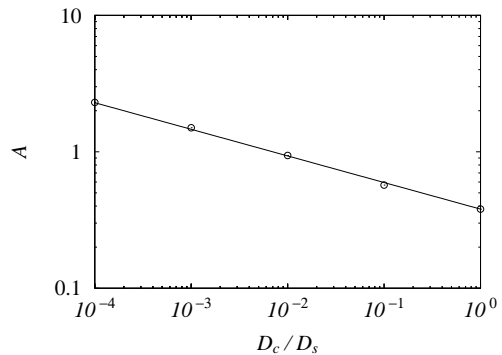}
\end{center} 
\caption{The proportionality coefficient $A$ as a function of the
corner diffusion constant $D_c$,
 which is equal to the edge diffusion constant $D_e$.
 The line represents the curve $A=0.38(D_c/D_s)^{-0.195}$.
}
\label{fig5}
\end{figure}

From Fig.\ref{fig3}(c) to (g), 
the island density $n$ seems to satisfy a new scaling relation
to the surface diffusion constant $D_s/F$
as
$n \propto (D_s/F)^{-\chi}$
with an exponent $\chi \approx 0.43$, 
when the density $n$ is lower than that of fractal islands.
In Fig.\ref{fig3} we depict fitting lines with an exponent $\chi= 3/7$,
which is expected from the critical mobile island size $i^*=3$ 
and from eq.(\ref{eq2}).
By fitting the island density in the scaling form with an exponent 3/7 as
\begin{equation}
n= A(\frac{D_s}{F})^{-3/7}
\label{eq3}
\end{equation}
the coefficient $A$ depends on the peripheral diffusion constant $D_c=D_e$.
By plotting $A$ versus $D_c/D_s$ in double logarithmic way as in Fig.\ref{fig5}
the coefficient $A$ is found to be proportional to $(D_c/D_s)^{-0.195}$.

\begin{figure}[h]
\begin{center} 
\includegraphics[width=0.45\linewidth]{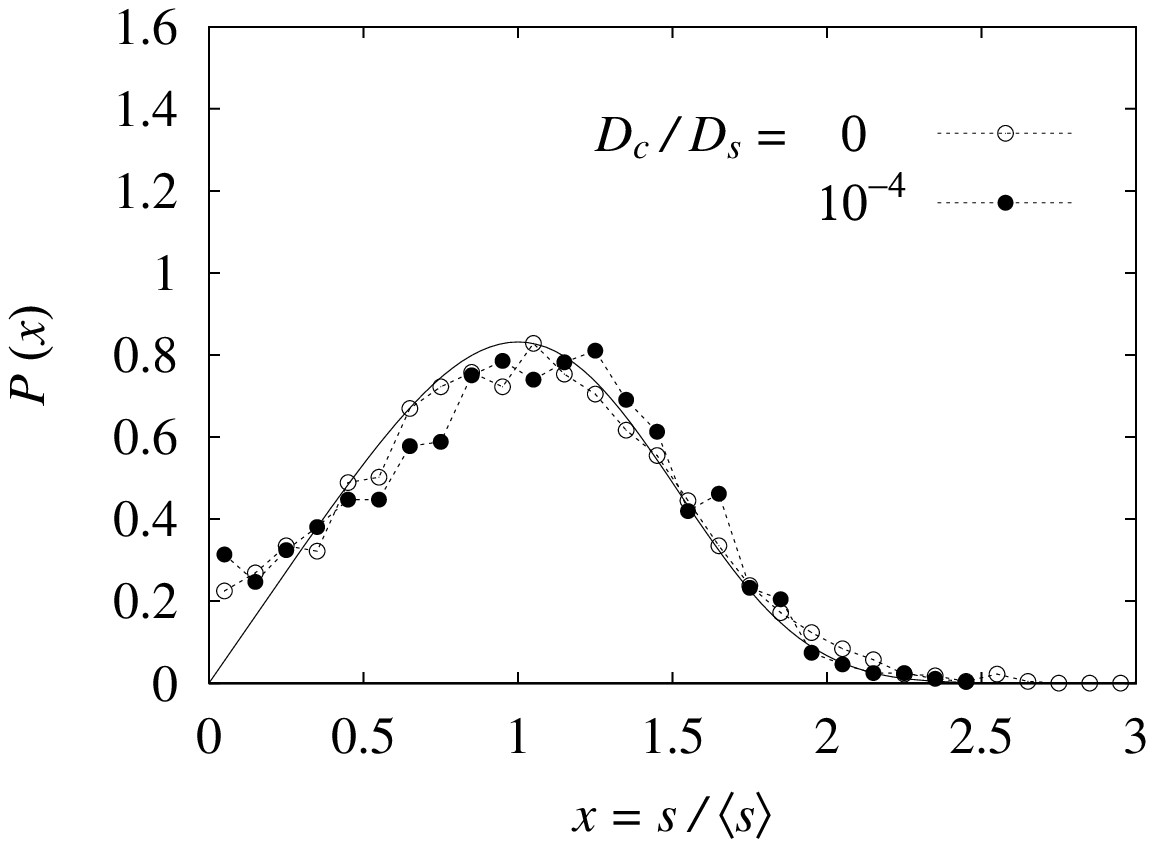}
\includegraphics[width=0.45\linewidth]{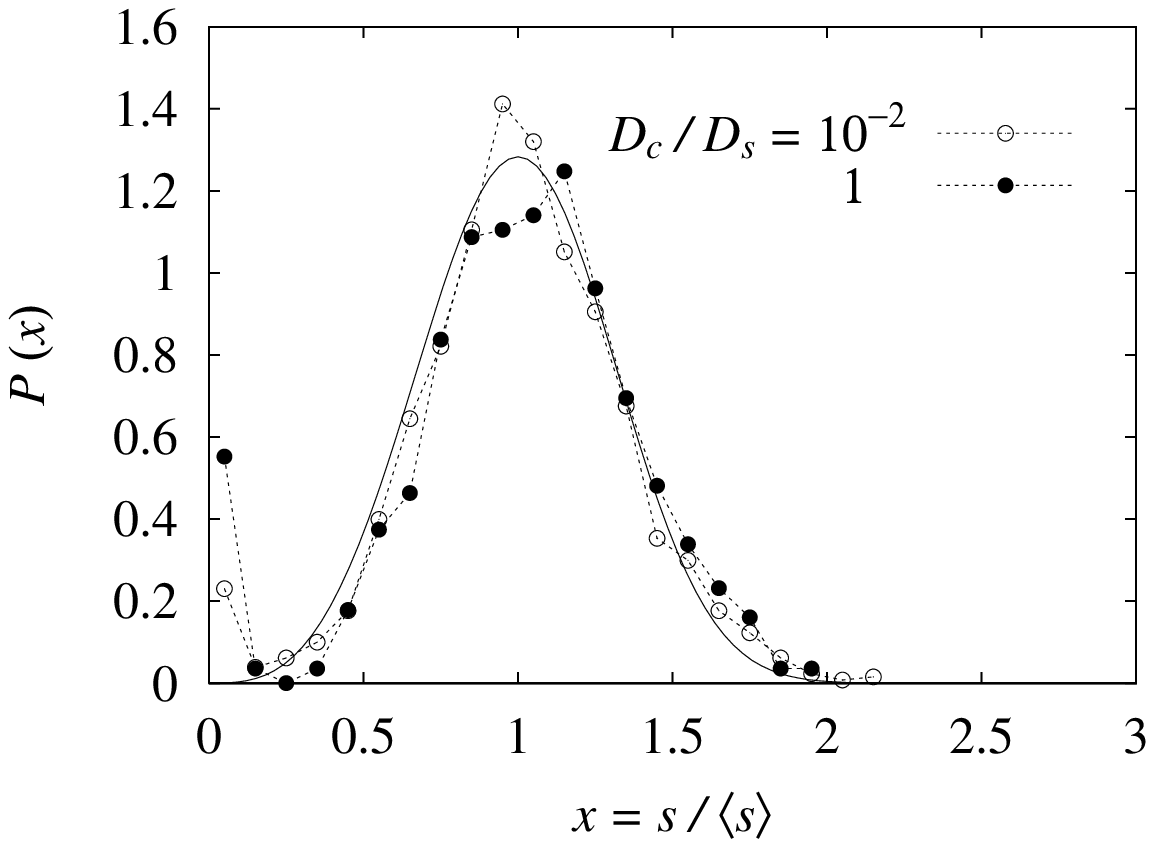}\\
(a) \hspace{5cm} (b)
\end{center} 
\caption{
Island size distribution with (a) low and (b) high corner diffusion constant,
$D_c$.
}
\label{fig6}
\end{figure}

From Fig.\ref{fig4}, one further notices that as the corner diffusion
increases,
not only the total number of islands but also the island size distribution
changes. 
\cite{bartelt+96,amar+95}
This tendency is apparent by plotting island size distribution
in Fig.\ref{fig6}.
Among the total $N_i=n L^2$ islands, there are $N(s)$ islands of a size $s$.
Since the average island size $\langle s \rangle$ changes as $D_c$ varies, 
the distribution of islands with a normalized size $x= s/ \langle s \rangle$ 
has the relevant concern.
We define the probability $P(x) \Delta x$ as the number of islands
with sizes between $x$ and $x+ \Delta x$ devided by the total number of islands
$N_i$ with $\Delta x=0.1$.
For each $D_c$, size distribution of 10 samples are averaged.
It is normalized as $\int_0^{\infty} dx P(x)=\int_0^{\infty} dx x P(x) =1$,
and corresponds to the scaled number density as
$N(s)=(\theta /\langle s \rangle ^{2}) P(s/\langle s \rangle )$
at a coverage $\theta$.
\cite{bartelt+96,amar+95}
At low $D_c$'s as $D_c=0$ and $10^{-4}D_s$ in Fig.\ref{fig6}(a),
island sizes are widely spread with a broad peak such
that there are very large islands with sizes 
about 2.5 times the average size.
This is the typical size distribution for $i^*=1$.
\cite{bartelt+96,amar+95}
On the contrary, at high $D_c$'s the island size distribution has a 
narrow peak around the average size, and extends
 less than twice the average size in 
the region of large sizes.
This is the typical size distribution for $i^* > 1$.
\cite{bartelt+96,amar+95,furman+97}
When the clusters smaller then $i^*$ are dissociable, 
the island size distribution is proposed\cite{amar+95} to
have the form
\begin{equation}
P_{i^*}(x)= C_{i^*} x^{i^*} \exp(-i^* a_{i^*} x^{1/a_{i^*}})
\end{equation}
with 
\begin{equation}
\frac{\Gamma[(i^*+2) a_{i^*}]}{\Gamma[(i^*+1) a_{i^*}]}
=(i^* a_{i^*})^{a_{i^*}}, \quad \mbox{and} \quad 
C_{i^*} = \frac{(i^* a_{i^*})^{(i^*+1)a_{i^*}}}{a_{i^*}\Gamma[(i^*+1) a_{i^*}]}
,
\end{equation}
where $\Gamma$ means the Gamma function.
The  probabilities corresponding for $i^*=1$ and $i^*=3$ 
are drawn  in 
Fig.\ref{fig6}(a) and (b) by continuous curves, respectively,
and fit quite well with the simulation data,
even though the dissociation is not allowed there.
In fact, the similar size distributions with mobile clusters
are obtained in the previous
Monte Carlo simulation.
\cite{furman+97}
There, however, no peripheral diffusion is granted, and the
grown islands had fractal structures, in constrast to our square
shaped clusters.

\section{Discussions and conclusions}

We have seen that a large corner diffusion influences 
the island number density 
and the size distribution drastically.
Without or weak corner diffusion, the island density varies in proportion to
$(D_s/F)^{- \chi}$ with an exponent $\chi=$1/3, 
irrespective of the island morphology.
This means that the nuclei are formed at the initial 
stage of deposition, as soon as the two adatoms collide with each other.
They are immobile and act as a center of nucleation.
Therefore, the randomness of the two-adatom encounter is frozen in.
After sufficient density of nuclei is formed, 
further deposited adatoms are incorporated into the preexisting
clusters. They develope into various shapes, 
depending on the edge and corner diffusion constants. 
Since the nucleation takes place at random, the island
sizes are distributed rather broad, reflecting the randomness
in the island separation. In any case,
this senario permits the island density to be independent of the morphology:
The island density is determined long before the island shape appears.

With a large corner diffusion, on the other hand, inspections of the
time evolution reveal that clusters of small sizes $i \le 3=i^*$ 
become mobile.
Since atoms in a cluster smaller than the critical size $i^*=3$ can move
around each other via the corner diffusion,
small clusters migrate around randomly on the substrate surface.
During this cluster diffusion, they coagulate with isolated atoms 
and as a result 
the island density decreases. 
Of course, islands larger than the size $i^*=$3 can migrate in some configurations,
but when compact islands 
with doubly or more bonded atoms are formed,
they cease migration on the substrate surface,
and start to act as a nucleation center.
During the further stage of cluster growth, the adsorbed atoms
form small mobile clusters  and are incorporated into islands in group.

This situation is treated by Villain et al. \cite{villain+92a}
in terms of
the rate equation, and
the island density $n$ is expected to follow the scaling relation
eq.(\ref{eq2}).
Since the trimer can migrate on the surface, $i^*=3$, 
$n \propto F^{3/7}$ is expected.
This is the reason why we fitted the density with a scaling exponent 
$\chi=3/7$ in Fig.\ref{fig3}.
This explains the $F$-dependence of $n$, but as for the dependence
on diffusion constants, we have to know $D_2$ and $D_3$ in terms of the
present edge and corner diffusion constants, $D_e$ and $D_c$.

There is an exact calculation by Sanchez et al.
\cite{sanchez+99} of diffusion constants of small clusters mediated by a
peripheral diffusion. The result is simple for a dimer and a trimer on
a square lattice as
\begin{equation}
D_2=\frac{D_c}{2}, \qquad D_3=\frac{D_e D_c}{3(D_e+D_c)}.
\end{equation}
For a dimer to diffuse, one of the adatoms which compose dimer
 has to cross round the corner.
For a trimer, for example, in an extended form to move 
an edge atom has to cross the corner and to slide on the edge to
another corner.
Without edge diffusion, the center of mass of a trimer cannot move.
In the case shown in Fig.\ref{fig3}, $D_e=D_c$ and thus $D_3=D_c/6$.
Inserting these exact results into the expression (\ref{eq2})
obtained by rate equation,
the island number density is expected to follow the relation
\begin{equation}
n \propto \left( \frac{F^3}{D_sD_c^2} \right)^{1/7}
\end{equation}
If this scaling holds, the coefficient $A$ in eq.(\ref{eq3}) should
be proportional to $A \propto (D_c/D_s)^{-2/7} = (D_c/D_s)^{-0.286}$,
instead the exponent -0.195 found in Fig.\ref{fig5}.
Therefore, the dependence of $n$ on cluster diffusion constant
does not seem to be properly given by the rate equation.

To check the scaling behavior of the island density with the crossover
from the monomer- to the trimer-limited form,
let us study the scaled island density $\tilde n= n(D_s/F)^{1/3}$.
It should be constant around 0.23 for small edge and corner diffusion
constnat $D_e=D_c$. For large $D_e=D_c$, it should be proportional to 
 $( D_s^2F/D_c^3)^{2/21}$, according to the result of the rate
 equation analysis (\ref{eq2}).
Therefore, we try the plot $\tilde n$ versus $\tilde F_1 = D_s^2F/D_c^3$, but
the data does not collapse on a single universal curve.

\begin{figure}
\begin{center} 
\includegraphics[width=0.8\linewidth]{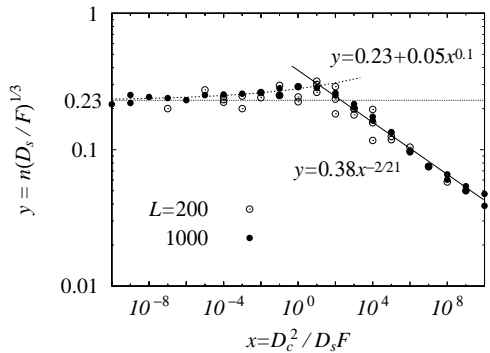}
\end{center} 
\caption{
The scaled island density $y= n (D_s/F)^{1/3}$ as a function of the 
scaled corner diffusion constant $ x= D_c^2 / D_s F$.
The dashed curve represents small $x$ asymptotic, $y=0.23+0.05x^{0.1}$,
and the continuous line the large $x$ asymptotic form $y=0.38x^{-2/21}$.
}
\label{fig7}
\end{figure}

From our simulation result, the island density behaves as
$n=A (D_s/F)^{-3/7}$ with an $D_c$-dependent coefficient 
$A = 0.38(D_c/D_s)^{-0.195}$
 for large $D_c$, as depicted in Fig.\ref{fig4}.
Since the exponent 0.195 is approximately equal to 4/21,
the density is expected to be scaled by the 
variable $x= D_c^2/ F D_s$ at a large $D_c$ more properly.
The scaling plot is shown in Fig.\ref{fig7}
for a wide range of combinations of diffusion constants
as $D_s/F=10^5 \sim 10^{11}$ and $D_e= D_c= 10^{-10}D_s \sim D_s$, 
and one observes a nice data collapse on a universal curve. 
It is fitted as
\begin{equation}
n \left( \frac{D_s}{F} \right)^{1/3} = \left\{
\begin{array} {ll}
0.23 +0.05 x^{0.1}  & \mbox{for $x \rightarrow 0$}\\
0.38 x^{-2/21} & \mbox{for $x \rightarrow \infty$} ,
\end{array} \right.
\end{equation}
as shown in Fig.\ref{fig7}.

The effect of the edge and corner diffusion during the homoepitaxial
nucleation growth on a singular surface is now summarized as follows.
As for the island morphology, these peripheral diffusion makes
the island regular and compact.
With only the surface diffusion, the irreversibly nucleated islands
have a fractal shape characteristic to DLA.
With the edge diffusion, islands become thick and compact.
The corner diffusion affects the shape anisotropy. 
One may typically say that without the corner diffusion dendritic 
needle  pointing in the $<10>$ direction appear,
 whereas with it squares with \{10\} faces appear.

The density of islands is mainly determined before the shape appears.
Without or with a weak corner diffusion, the critical island size is
$i^*=1$, and the density $n$ is related to the surface diffusion constant $D_s$
normalized by the deposition rate $F$ 
as $n \propto (D_s/F)^{-1/3}$, irrespective of island morphology.
With a large corner diffusion constant, on the other hand, 
small clusters $i \le 3$ can migrate randomly on the substrate surface, and
another scaling relation 
$n \propto F^{3/7}/ D_s^{5/21} D_c ^{4/21} = (F/D_s)^{3/7} (D_s/D_c)^{4/21}$ 
is found appropriate.
The flux $F$-dependence agrees with the rate equation theory with $i^*=3$,
but the dependence on diffusion constants $D_c$ contradicts 
the theory.  The reason of this discrepancy is not clear.
But, since the rate equation is a mean-field type approximation
which neglects spatial correlation, it may be fortuitous that $F$ dependence
agrees in the simulation and the rate equation theory.

Recently, the mobility of large clusters is  proposed to be very effective
in selecting the island size and in arranging islands in order 
in heteroepitaxial growth.\cite{liu+01}
Our model allows the motion of only small clusters less than the size $i^*=3$
by the corner diffusion, and some size selection is observed.
Due to the absence of any direct interaction between islands,
however, the ordering in island positions cannot be achieved.
Our model neither contains the size-limiting effect,
and the coarsening takes place as the corner diffusion increases.

\acknowledgements
This work is supported by Grant-in-Aid for Scientific Research 
from Japan Society of the Promotion of Science.
The author is benefited from the inter-university
cooperative research program of the Institute for
Materials Research, Tohoku University.


\end{document}